\documentclass{JHEP}
\usepackage{amsmath}
\usepackage{amssymb,amsfonts}
\usepackage{epsfig}

\newcommand{\ie}{{\em i.e. }}
\newcommand{\Cs}{\mathbb{C}^*}

\newcommand{\CP}{\mathbb{C}\mathbb{P}}
\newcommand{\RP}{\mathbb{R}\mathbb{P}}
\newcommand{\Z}{\mathbb{Z}}
\newcommand{\C}{{\cal C}}
\newcommand{\R}{\mathbb{R}}
\newcommand{\G}{{\cal G}}

\newcommand{\N}{{\cal N}}

\newcommand{\Ka}{K{\"a}hler }

\newcommand{\Co}{\mathbb{C}}
\newcommand{\I}{{\cal I}}

\newcommand{\pamatrix}[1]{\begin{pmatrix} #1 \end{pmatrix}}
 
\newcommand\fverb{\setbox\pippobox=\hbox\bgroup\verb}
\newcommand\fverbdo{\egroup\medskip\noindent%
                        \fbox{\unhbox\pippobox}\ }
\newcommand\fverbit{\egroup\item[\fbox{\unhbox\pippobox}]}
\newbox\pippobox
\title{Phases of Supersymmetric 
Gauge Theories from M-theory 
on $G_2$ manifolds}

\author{P. Kaste$^1$, A. Kehagias$^2$ and H. Partouche$^1$ \\
$^1$ Centre de Physique Th{\'e}orique, Ecole Polytechnique\\
F-91128 Palaiseau cedex, FRANCE\\ 
 E-mail: \email{Peter.Kaste, Herve.Partouche@cpht.polytechnique.fr}\\
\vskip .2 in 
$^2$ Physics Department, NTUA, 15773 Zografou, Athens, GREECE\\
        E-mail: \email{kehagias@mail.cern.ch}}
\preprint{\hepth{0104124}\\ CPHT-S019.0401}

\abstract{We consider M-theory on compact spaces of $G_2$ holonomy constructed 
as orbifolds of the form (CY$\times S^1)/\Z_2$ with fixed point set 
$\Sigma$ on the CY. This describes ${\cal N}=1$ $SU(2)$
gauge theories with $b_1(\Sigma)$ chiral multiplets in the
adjoint. For $b_1=0$, it generalizes to compact manifolds the study of
the phase transition from the non-Abelian to the confining phase
through geometrical $S^3$ flops.  
For $b_1=1$, the non-Abelian and Coulomb  phases are  realized, where 
the latter arises by desingularization of the fixed point set, while
an $S^2\times S^1$ flop occurs.
In addition, an extremal transition between  $G_2$ spaces can 
take place at conifold
points of the CY moduli space where unoriented membranes wrapped on
$\CP^1$ and 
$\RP^2$ become  massless.  
}


\begin{document} 

\maketitle %%%%%%%%%% THIS IS IGNORED %%%%%%%%%%%

\section{Introduction}

There are many ways to obtain ${\cal N}=1$ theories in 4D. The initial
example is the heterotic compactification on 
Calabi-Yau threefolds. The  $SU(3)$ holonomy of the later allows a
surviving Killing spinor and leads to an ${\cal N}=1$ gauge theory
in 4D. Similarly, a compactification of M-theory on a 7-manifold
of $G_2$-holonomy  gives rise to an ${\cal N}=1$ theory in 4D. However,
since there are no gauge fields to start with in
11D, the vector multiplets are generically Abelian and thus of restricted
interest. Their number depends on the
topological properties of the internal space and equals the number
of non-contractible two-cycles.  On the
other hand, it is known that the appearance  of singularities in
the internal space leads to non-Abelian gauge theories. As has
been shown in \cite{Sen:1997kz}, in certain cases, as  two-cycles are
collapsing, $A_n$ singularities  appear and a corresponding
enhancement of the gauge group occurs. This has explicitly been shown for
the KK monopole when the ALE space  degenerates to
 $\mathbb{R}^4/\mathbb{Z}_n$. In this limit, there
exists an enhancement of the gauge group to $SU(n)$ due to
 membranes wrapped on intersecting two-cycles.  This is the way
non-Abelian vectors can appear in M-theory. This mechanism was also considered
 in \cite{FKPZ} in order to study ${\cal N}=1$ 6D theories at the
conformal point in the AdS/CFT correspondence.

M-theory compactifications on $G_2$-holonomy manifolds have
appeared in the past \cite{Papadopoulos:1995da}. Recently, phase
transitions between 
different $G_2$ manifolds have been considered in \cite{Partouche:2001uq}
that describe changes of branches in the scalar potential or a 
Higgs mechanism in the Abelian sector of the theory. 
Moreover, the possibility of understanding
the non-Abelian phase structure of M-theory on $G_2$ manifolds was
pointed out in \cite{Acharya:2000gb,Atiyah:2000zz,Acharya:2001dz}. 
The central ingredient in \cite{Atiyah:2000zz}
is an $S^3$-flop in the underlying geometry. The flop has been
interpreted as a phase transition of the gauge theory. On one side the 
$SU(N)$ gauge group is in its non-Abelian phase,
 while on the other side it has
disappeared as an effect of confinement.  However, the
$G_2$ space which has been employed in this construction was 
non-compact. 

In the present work we describe these effects in the case of compact spaces 
of $G_2$-holonomy constructed as orbifolds of the form (CY$\times
S^1)/\Z_2$. The $\Z_2$ acts  as an
inversion on the $S^1$ coordinate $x^{10}$ and antiholomorphically on the CY, 
 so that $J\to -J$ and $\Omega\to \bar \Omega$, where 
$J$, $\Omega$ are the \Ka form and the holomorphic 3-form. As a result,  
\begin{equation}
\Phi=J\wedge dx^{10}+\Re(\Omega) 
\label{g2struct}
\end{equation}
provides the orbifold with a $G_2$ structure \cite{Joyce2}. 
The fixed point 
set $\Sigma$ of the antiholomorphic involution on the CY are special Lagrangian
3-cycles. When it is composed of $P$ disconnected components
$\Sigma_p$ $(p=1,...,P)$, each of them  
is promoted to an associative 3-cycle of $A_1$ singularities on
the 7-space. In M-theory compactifications, the first Betti number
$b_1(\Sigma_p)$ of this cycle then counts the number
of chiral multiplets in the adjoint representation of an 
$SU(2)$ gauge group.
The case $b_1(\Sigma_p)=0$, to be considered in section \ref{without}
generalizes the work of  
\cite{Atiyah:2000zz} to a $SU(2)$ gauge group on compact
$G_2$ spaces. When $b_1(\Sigma_p)=1$, we will see in section \ref{MM}
that the non-Abelian and Coulomb
 phases also occur in M-theory. The Coulomb phase will be described
geometrically as the desingularization of the orbifold fixed points. In
addition,  $S^2\times S^1$ flops will take place.  

From the non-Abelian point of view, the transitions in the
two models with $b_1(\Sigma)=0$ that
we will present, occur at different types of singular points in the CY 
complex structure moduli space.   In one case the CY has ordinary
nodes and a conifold transition \cite{Candelas:1990ug}
gives rise to an additional
descendant $G_2$ branch. This will generalize the other non-compact $G_2$
manifold with $A_1$ singularities appearing in the literature 
\cite{Partouche:2001uq}. These transitions are triggered by 
black holes \cite{Strominger:1995cz,Greene:1995hu} described as 
{\em unoriented} membranes wrapped 
on non-calibrated 
$\CP^1$ and $\RP^2$.  A discussion along these lines
from the supergravity point of view can be found in \cite{Pap}.

We remark that these models are dual to type IIA compactified on the CY 
taking part in the orbifold in the presence of $2b_1(\Sigma_p)$ D6-branes 
and $b_1(\Sigma_p)$ O6-planes wrapped on $\Sigma_p$ 
\cite{Gomis:2001vk,Edelstein:2001pu,Kachru:2001je,Cachazo:2001,
Edelstein:2001b}.

%%%%%%%%%%%%%%%%%%%%%%%%%%%%%%%%%%%%%%%%%%%%%%%%%%%%%%%%%%%%%%%%%%%%%%%%%%%%%

\section{Models without adjoint matter}
\label{without}

Our aim is to construct $G_2$ manifolds out of products CY$\times
S^1$ on which we act by antiholomorphic involutions $w$ 
on the CY and by inversion $\I$ on $S^1$. In
\cite{Partouche:2001uq}, freely acting involutions $w$ on the 
whole CY ambient space were considered that result in smooth
7-dimensional orbifolds. Upon compactifying M-theory on them,
transitions between topologically different such manifolds were used
to describe transitions between distinct branches in the scalar potential of
the associated $\N=1$ Abelian effective field theory. 
However, it was noticed that if the
involution has fixed points in the CY ambient space, 
a subset of them can live on the 7-dimensional $G_2$-orbifold 
and may or may not be desingularized while keeping the holonomy within
$G_2$. 
The
fixed point set on the CY is then a special Lagrangian 3-cycle whose
topology can vary when changing the CY complex structure. As a result
it was claimed that a single CY moduli space can be split
into distinct components describing $G_2$ orbifolds with different
fixed point set topologies. 
When these orbifolds can be desingularized, they give
manifolds with different Betti numbers due to different ``twisted
sectors'' as will be seen in section \ref{MM}. 
In M-theory compactifications they give the various phases of the
$\N=1$ non-Abelian effective field theory.

For the moment, we would like to consider the 
simplest case, where a transition
takes place between a phase without fixed points to a phase with fixed
point topology of disjoint 3-spheres. This transition was actually already
considered in \cite{Atiyah:2000zz} and \cite{Partouche:2001uq} in a 
local version on non-compact $G_2$ manifolds, namely, the spin bundle over
$S^3$ and $(T^*S^3\times \R)/\Z_2$, respectively. They both describe  phase 
transitions in pure $SU(2)$ gauge theory. However,  the model 
in \cite{Partouche:2001uq} has an additional $G_2$ branch arising from 
a conifold transition of the underlying CY $T^*S^3$. 
These effects are not restricted to non-compact $G_2$ spaces and 
this section is dedicated to demonstrate them in a compact model.
In order to keep the discussion as simple as possible, we consider
$G_2$ manifolds based on CY hypersurfaces in (the resolution of) the
well-known weighted projective space $\CP^4_{11222}$.

\subsection{A model based on {$(\CP^4_{11222}[8] \times S^1)/ \Z_2$}}
\label{model1}

We define the one-parameter sub-set of threefolds $\C_1$ in the
family 
$\CP^4_{11222}[8]$ with
Hodge numbers $h_{11}=2$, $h_{12}=86$ by
\begin{equation}
p_1 = z_6^4(z_1^8 + z_2^8-2\phi z_1^4z_2^4)+z_3^4+z_4^4+z_5^4=0\; ,
\label{def} 
\end{equation}
where the projective coordinates are subject to the identifications
under the two $\Cs$ actions
\begin{equation}
\begin{tabular}{c|cccccc}
 & $z_1$ & $z_2$& $z_3$&$ z_4$& $z_5$&$ z_6$ \\ \hline
$\Cs_1$& $0$&$0$&$1$&$1$&$1$&$1$ \\
$\Cs_2$& $1$&$1$&$0$&$0$&$0$&$-2$ 
\end{tabular}\; . \label{scalings}
\end{equation}
The additional variable $z_6$ accompanied by a second scaling action
arises from blowing up the $\Z_2$ singularity $\{z_1=z_2=0\}$ of
$\CP^4_{11222}$ which the CY family intersects. 
The resulting toric ambient space has an excluded set  
\begin{equation}
(z_1,z_2)\neq (0,0) \quad \mbox{and}\quad
(z_3,z_4,z_5,z_6)\neq(0,0,0,0)\; .
\label{forbid}
\end{equation}
Notice that among the large number of monomials allowed by the
$\Cs$ actions, we choose to consider only one, namely 
$z_1^4 z_2^4 z_6^4$, in Eq. (\ref{def}) with arbitrary complex coefficient. 

A CY in this family happens  to be singular when the equation
(\ref{def}) is non-transverse, \ie\ when $p=\partial_{z_i} p=0$, $(i=1,...6)$. 
This occurs only when $\phi=\pm 1$ and for 
$\{z_1^4=\phi z_2^4\}\cap \{z_3=z_4=z_5=0\}$.
Thanks to Eq. (\ref{forbid}) these singularities lie in charts where
$z_2$ and $z_6$ are non vanishing, so that we can
rescale them to unity. In each case, there are thus 4 singular points,
\begin{equation}
\begin{array}{llll}
\mbox{for }\phi=+1 &~:~ & (i^k,1,0,0,0,1)& ,(k=0,...,3)\\   
\mbox{for }\phi=-1 &~:~ & (i^k e^{i\pi/4},1,0,0,0,1) &,(k=0,...,3)\; .
\label{nodes}
\end{array}
\end{equation}
At these points, the determinant of second
derivatives $\det(\partial_A\partial_Bp_1)$ $(A,B=1,3,4,5)$ {\em does vanish},
 so that these isolated singularities
{\em are not} nodal points. 
Therefore,
if a transition  to another CY occurs at $\phi=\pm1$, it is not
of the usual conifold type.   
Note that the holomorphic change 
$z_1 \to e ^{ i\pi/4} z_1$  leaves $p_1$
invariant if we substitute simultaneously $\phi \to
-\phi$. Therefore, from the CY point of view,
 $\phi$ and $-\phi$ parametrize the same complex
structure and we could restrict to $\Re(\phi) \geq 0$.

However, we want to construct
$G_2$ orbifolds of the form
\begin{equation}
\G_1=(\C_1\times S^1)/\sigma\; 
\end{equation}
with an 
involution $\sigma=w\I$, where $\I$ acts as an inversion on $S^1$ and
$w$ is antiholomorphic with fixed points on the CY ambient space  
\begin{equation} 
\sigma \; : \quad z_i\to \bar{z}_i\quad (i=1,...,6)\quad , \qquad x^{10}\to
-x^{10}\; , \label{s} 
\end{equation}
where $x^{10}$ is the $S^1$ coordinate. Clearly $w$ commutes with
the $\Cs$ actions. For  $\G_1$ to be well defined, $\phi$ should be real so
that $w$ is a symmetry of the CY. Note that
 $\phi$ and $-\phi$ are no longer equivalent for $\G_1$
since $z_1 \to e ^{ i\pi/4} z_1$
does not commute with $\sigma$. 
The single family of CY's for arbitrary
complex $\phi$ thus splits {\em a priori} into three $G_2$ 
branches for the orbifold
parametrized by real $\phi$,
namely $\phi <-1$, $-1<\phi<1$ and $\phi >1$.    

Let us determine now the fixed point set of the orbifold. If we denote
by $\Sigma$ the special Lagrangian 3-cycle in $\C_1$ fixed by $w$, the
total fixed point set is just two copies of $\Sigma$, one at $x^{10}=0$
and the second at $x^{10}=\pi R$, where $R$ is the radius of $S^1$. A
point $M=[z_1,\ldots,z_6]$ then belongs to $\Sigma$ if it solves
(\ref{def}) and if its equivalence class is the same as
$[\bar{z_1},\ldots,\bar{z_6}]$. This means that there exist
$\lambda_1,\lambda_2\in \Cs$, such that 
$z_i=\rho_i \bar{z_i}$, $(i=1,\ldots,6)$ with 
$\rho_i=\prod_{j=1}^2 \lambda_j^{Q^{(j)}_i}$, where $Q^{(j)}_i$ is the
weight of $z_i$ under $\Cs_j$.
The scalings (\ref{scalings}) and constraints (\ref{forbid}) imply
that $|\rho_i |=1$, so that $\rho_i ^{-1}=\bar{\rho_i}$. Therefore,
if we define $z'_i=\rho_i ^{-1/2} z_i=x_i+iy_i$ $(i=1,\ldots,6)$, we
can write $M=[z'_1,\ldots,z'_6]$ with $z'_i=\bar{z}'_i$.
Without loss of generality,
$\Sigma$ is therefore determined by the defining equation (\ref{def}) for
real unknowns $x_i$, 
\begin{equation}
x_6^4(x_1^8+x_2^8-2\phi x_1^4x_2^4)+x_3^4+x_4^4+x_5^4=0 \; . 
\label{def2}
\end{equation}
Notice that if $x_6$ could vanish in this equation, it would imply
$x_{3,4,5}=0$ as well, which is forbidden. Hence we can rescale
$x_6$ to 1. The same is also true for $x_2$, since a  vanishing $x_2$
would also imply $x_{1}=0$. The scaling actions being
gauged away,  Eq. (\ref{def2}) is solved for $x_1^4$, 
\begin{equation}
x_1^4 =\phi \pm\sqrt{\phi^2-(1+x_3^4+x_4^4+x_5^4)}\; . 
\label{def3}
\end{equation}
We see that for $\phi<1$, there is no real solution and thus $\Sigma$
is empty. For $\phi=1$, when the CY is singular, there are solutions
for $x_3=x_4=x_5=0$, $x_1^4=1$ and $\Sigma$ consists of two points
\begin{equation}
\Sigma=\{(1,1,0,0,0,1),(-1,1,0,0,0,1)\}\, .
\end{equation}
Finally, for $\phi>1$, we find it convenient to define the variables
\begin{equation}
u_1= x_1^4-\phi \;\; , \quad u_j=x_j^2 \; \mbox{sign}(x_j)\; , \quad
(j=3,4,5)\; ,
\end{equation}
in which the equation for $\Sigma$ reads 
\begin{equation}
u_1^2+u_3^2+u_4^2+u_5^2=\phi^2-1\; , \label{Sph} 
\end{equation}
describing an $S^3$ of radius $\sqrt{\phi^2-1}$. 
However, whereas $u_{3,4,5}$ and $x_{3,4,5}$ are in one-to-one
correspondence, the map from $x_1$ to $u_1$ is two-to-one. Furthermore, 
from  
$-\sqrt{\phi^2-1}\leq x_1^4-\phi\leq\sqrt{\phi^2-1}$
we get 
\begin{equation}
\begin{array}{lll}
&&0<(\phi-\sqrt{\phi^2-1})^{1/4}
\leq x_1\leq (\phi+\sqrt{\phi^2-1})^{1/4}\; \\ &\mbox{ or }&
\phantom{0}-(\phi+\sqrt{\phi^2-1})^{1/4}
\leq x_1\leq -(\phi-\sqrt{\phi^2-1})^{1/4}<0\; , 
\end{array}
\label{xx}
\end{equation}
so that $\Sigma$ actually consists of two copies of $S^3$ that do not
intersect, one with $x_1>0$ and the other with $x_1<0$. As a result,  
the fixed point set $\Sigma$, according to the values of $\phi$ is
\begin{eqnarray}
\begin{array}{ll}
\phi<1,~~~ & \Sigma={\emptyset} \quad\mbox{(no fixed
  points)}\; , \\
\phi=1,~~~ & \Sigma=\{(\pm 1,1,0,0,0,1)\}\; ,\\
\phi>1, ~~~& \Sigma= S^3\cup S^3 \quad \mbox{of radii }\sqrt{\phi^2-1}\; .
\end{array} \label{ss1}
\end{eqnarray}
We thus have two distinct phases as $\phi$ varies. 
Also, we see that the two special Lagrangian 3-spheres present for $\phi>1$
shrink to the isolated singularities given in Eq. (\ref{nodes}) for
$k=0,2$ when $\phi\to 1$.

\vskip .2in

\noindent
{\it Spectrum for $\phi<1$}
\vskip .05in 

Let us now consider M-theory compactified on the previous orbifolds. 
When $\phi<1$, the compact
space is smooth and the four-dimensional massless spectrum of
M-theory is then determined by the Betti numbers of $\G_1$
\cite{Papadopoulos:1995da}. The $b_3$ deformations of the $G_2$ structure
$\Phi$ together with the flux of the eleven dimensional supergravity
3-form potential $C$ on the 3-cycle homology classes give rise to
$b_3$ complex scalars. In addition, the
dimensional reduction of $C$ on the 2-cycle homology classes
provides us with $b_2$ vector bosons in four dimensions. Together with
$\N=1$ superpartners from the reduction of the eleven dimensional
gravitino, we have $b_2$ $\N=1$ vector multiplets and $b_3$ neutral
chiral multiplets. In our case, a 2-cycle on $\G_1$ arises from a
2-cycle in $\C_1$ even under $w$, while the odd ones times $S^1$ give
invariant 3-cycles on $\G_1$. Also, 3-cycles in $H_{1,2}$ ($H_{0,3}$) and
$H_{2,1}$ ($H_{3,0}$) combine to give an equal number of even and odd
3-cycles.   
If we denote by $h_{11}^\pm$ the number of
even and odd homology classes of 2-cycles on $\C_1$, then the 
relevant Betti
numbers of the smooth $G_2$ orbifold $\G_1$ for $\phi<1$ are
\begin{equation}
b_2=h_{11}^+ \quad , \quad b_3= \frac{h_{30}+h_{03}}{2} +
\frac{h_{21}+h_{12}}{2}  +h_{11}^-=1+h_{12}+h_{11}^-\;.
\label{b23}
\end{equation}
In our case there are $h_{11}=2$ cohomology classes on $\C_1$. The
first one is the pullback of the \Ka form $J$ that determines the size
of the ambient $\CP^1_{11222}$ on which $w$ acts antiholomorphically as
a symmetry, implying $J\to -J$. Similarly, the second cohomology
class proportional to $dz_6 \wedge d\bar{z}_6$ is also odd under
$w$. Therefore, $h_{11}^+=0$ and the Betti numbers read 
$$
b_2=0 \qquad \mbox{ and } \qquad b_3=1+86+2=89\; .
\label{q}
$$ 
The spectrum thus consists in
\begin{equation} 
\mbox{89 chiral multiplets .}
\end{equation}

\vskip .2in

\noindent
{\it Spectrum for $\phi>1$}
\vskip .05in 

In this case the massless spectrum decomposes into two pieces. The
``untwisted sector''
arises as before from the dimensional reduction of the eleven
dimensional supergravity multiplet on the 2- and 3-cycles even
under the orbifold involution $\sigma\equiv w\I$. As a result, it
still consists of $b_2=0$ $\N=1$ vector multiplets and $b_3=89$ chiral
multiplets. To this one has to add the states localized on
the fixed point set
$\Sigma\times \{0,\pi R\}$, \ie 4 disconnected copies 
of $S^3$, \cite{Acharya:2000gb}. 
Around each of these
singular points, the geometry looks like $\R^4/\Z_2\times \R^3$, where
3 directions in $\R^4$ correspond to the imaginary parts of local 
coordinates $Z_j$ $(j=1,2,3)$ that transform as
$Z_j\longrightarrow\bar{Z}_j$ under $w$, while the fourth direction
accounts for the $S^1$, while the factor $\R^3$ is the tangent plane on
each $S^3$. As the ``twisted states" are localized around each singular set
$S^3$
and that the latter are disjoint, 
we can study the spectrum arising from only one
of them. The total spectrum will consist of four copies of it. 
 Since we do not take
into account here the effect of an eventual generation of a
superpotential from M-theory instantons \cite{Harvey:1999a}, we can
determine the 
desired spectrum by use of an adiabatic argument that consists in
considering first the situation where $S^3$ is of large volume, \ie\
$\phi$ large and positive. In that case M-theory is actually
compactified down to 7 dimensions on a 4-space with an $A_1$
singularity. The resulting bosonic spectrum consists of a vector
field in the adjoint of $SU(2)$. The bosonic spectrum for finite
$\phi$ is then simply obtained by dimensional reduction of this vector
field on $S^3$ and gives an $SU(2)$ vector boson in 4 dimensions plus
$b_1(S^3)=0$ real scalars (no Wilson lines). $\N=1$ 
supersymmetry then assures that we have in total for $\Sigma\times \{0,\pi
R\}$  
\begin{equation}
\mbox{1 vector multiplet of }SU(2)^4 \qquad \mbox{and} \qquad 
\mbox{89 neutral chiral multiplets .}
\end{equation}
   
One can now ask if it is possible or not to desingularize the
orbifold fixed points and obtain a smooth manifold of $G_2$ holonomy. 
For each $S^3$,
this means blowing up at each point a
2-sphere  in the transverse
4-space. In other words, one would glue 4 copies of Eguchi-Hanson
spaces $X_4$ times $S^3$. The resulting smooth manifold would
then have 4 more moduli $v_{n}$ $(n=1,...,4)$ parametrizing the volumes of the
4 blow up $S^2$'s. However, on a $G_2$ manifold, there are only
3-cycle moduli and the $S^2$'s have to combine with a 1-cycle
$\gamma_1$ of radius $r$ so that the $G_2$ moduli would take the form
$v_{n}r$. Since $b_1(X_4)=0$, $\gamma_1$ should be chosen on
$S^3$. However, the latter has $b_1=0$ and the resolution is not possible.
From this discussion, we see that  a
 necessary condition for a $G_2$ resolution of
the orbifold singularities to be possible is
$b_1(\Sigma_p) \geq 1$, where $\Sigma = \cup_p^P \Sigma_p$, for connected 
3-cycles $\Sigma_p$. \footnote{ Notice that, partial 
resolutions should be possible 
when only some $\Sigma_p$'s have non-trivial $b_1$.}
 In fact, $\N=1$
supersymmetry confirms this. The geometrical moduli $v_{n}$
must be part of 3-cycle volumes so that they can be combined with 
the flux
of the eleven dimensional supergravity 3-form $C$ on the corresponding
 $S^2\times \gamma_1$ to give complex scalars of chiral multiplets.  
Notice that the condition $b_1(\Sigma_p) \geq 1$ is the necessary one
given by Joyce for a $G_2$ resolution to be possible and that, to be
sufficient, there must be in addition a nowhere vanishing harmonic
1-form on $\Sigma_p$ \cite{Joyce1}\footnote{Actually, in \cite{Joyce1} 
$\Sigma$ was
  implicitly supposed to be connected}. 
Unfortunately, this last condition happens to be 
difficult to check in practice \cite{Joyce11}.

We saw that the present model cannot be resolved while keeping the
holonomy in $G_2$. However, one can still wonder whether a Ricci 
flat resolution is possible, such that the holonomy is a subgroup of
$SO(7)$, but not in $G_2$. \footnote{{It is known
  \cite{Salamon:1989} 
that reduced holonomy $G_2$ for a metric on an oriented Riemannian
7-manifold implies Ricci flatness. }} 
If such Ricci flat resolutions would exist, they would
describe non-supersymmetric M-theory vacua. 
However, one can doubt that this is possible,  since
switching on the blow up moduli $v_{n}$ would smooth the compact
space, so that we pass to a Coulomb branch of $\N=1$ pure $SU(2)^4$  
super Yang-Mills theory with spontaneous breaking of
supersymmetry and we know that such a Coulomb branch does not exist.

\subsection{Physical interpretation of the $SU(2)^4$ phases}

The geometrical $G_2$ moduli space parametrizes the $G_2$ structure $\Phi$,
(\ref{g2struct}), on the underlying Riemannian 7-manifold. 
By varying this structure, we thus vary the volumes of the associative
3-cycles, which are calibrated w.r.t.\ $\Phi$, \ie their volume is
given by the integral of the pullback of $\Phi$.
When considering M-theory on $G_2$ orbifolds, these real moduli are
complexified by the flux of the M-theory 3-form $C$ through the
cycles and become the lowest component of chiral fields.

In the previous section
we distinguished domains according to the existence or
non-existence of fixed point locii of the antiholomorphic involution
(\ref{s}).  
When they exist, they are associative 3-cycles and via the $\Z_2$
singularity in their transverse space there is an $SU(2)$ gauge group
associated to each of them. The complexified (by the $C$-form flux)
volume $V_M$ of the associative 3-cycle becomes the complexified (by the
$\Theta$ angle) gauge coupling of the corresponding $SU(2)$ gauge group. 

Let's go back now to our model based on $\G_1$ to be more explicit. We
considered a one-dimensional subset of the geometrical $G_2$ moduli
space parametrized by the real modulus $\phi$. For $\phi >1$ we found
a fixed point locus consisting of four associative 3-spheres whose
volumes vanish as we send $\phi$ to unity. While varying $\phi$, the
real volumes of these four 3-cycles behave alike. 
Assuming that their homology class is actually the same so that also
the $C$-field fluxes coincide, the four associated gauge groups
have one and the same complex gauge coupling
\begin{equation}
V_M=  \frac{1}{g_{YM}^2}+i\Theta\, .
\end{equation}
For $\phi<1 $ the antiholomorphic involution acts freely and there is no
sign of any non-Abelian gauge symmetry.

The point in the M-theory compactification is \cite{Atiyah:2000zz},
that the physical moduli space is complex and that due to
holomorphicity of the $\N=1$ theory singularities occur at least in
complex codimension one. Hence, in M-theory we can continuously
deform from a theory at $\phi \gg 1$ to one at $\phi \ll -1$ without
encountering any singularity. The running of the coupling suggests that
the region $\phi \gg 1$ of large positive volume should correspond to
the UV, where the 
$SU(2)^4$ gauge theory is weakly coupled and nonperturbative
corrections are suppressed. 
M-theory in the region $\phi
\ll -1$ should then describe the confining phase of the same theory in
the IR, with no sign of gauge symmetry.

In the noncompact model of \cite{Atiyah:2000zz} the transition between
these phases could be interpreted as a flop transition
between two associative 3-spheres, such that only in one of the
geometric phases the orbifold group defining the theory acted freely.
We would now like to see if such a flop transition takes place in our
compact model as well. 

The two $S^3$'s in Eq. (\ref{ss1})
for $\phi>1$ being fixed point locii of the
antiholomorphic involution $w$ acting on the whole CY ambient space,
our strategy
is then to consider other involutions that would fix 
3-spheres for $\phi<1$ that vanish for $\phi \to 1^-$. 
If we consider first diagonal involutions 
$z_i\to e^{i\theta_i}\bar z_i$ $(i=1,...,6)$, 
one finds that there are actually 256
inequivalent choices of the phases $ \theta_i$ consistent with the defining
equation $p_1=0$. Among the 256 fixed point sets, 
16 of them present for $\phi>1$ vanish when  $\phi\to 1^+$ 
and turn out to be empty for $\phi<1$. 
Actually, these sets are  16 copies of
$S^3\cup S^3$. More precisely there exist 8 vanishing $S^3$'s for each of the
4 singular points in Eq. (\ref{nodes}) 
including the  ones that are fixed by the involution $w$ we chose.   

Another set of involutions consist   in  
$z_1\to e^{i\theta_1}\bar z_2, ~  z_2\to   e^{i\theta_2}\bar z_1, ~ z_j\to  
e^{i\theta_j}\bar z_j$ 
 $(j=3,4,5,6)$, where only  128 are  compatible with $p_1=0$. 
The fixed point sets of 8 of them happen to vanish at $\phi=1$ with finite
volume for $\phi<1$ and  the geometry of each one of them 
is determined by the equation
\begin{equation}
z_1^8+\bar z_1^8 +2\phi z_1^4\bar z_1^4+x_3^4+x_4^4+x_5^4 =0 \, , 
\end{equation}
where $z_1$ is complex and $x_{3,4,5}$ are real.
Defining $z_1^4=U_1+iV_1$, this gives
\begin{equation}
2 (\phi+1)U_1^2+x_3^4+x_4^4+x_5^4=2(1-\phi)V_1^2 \, , \label{tt}
\end{equation}
in which we can rescale $V_1=\pm 1$. 
We thus obtain, for each  of these 8 involutions 
\begin{eqnarray}
\begin{array}{ll}
\phi>1,~~~ & { \emptyset} \quad\mbox{(no fixed
  points)}\; , \\
\phi=1,~~~ & \{(i^k,1,0,0,0,1)\; , ~~~(k=0,1,2,3)\}\\
\phi<1, ~~~&  \bigcup_{n=1}^4 \tilde S^3 \quad \mbox{of radii }
\sqrt{2(1-\phi)}\; . 
\end{array}
\end{eqnarray}
%Hence, there is a total of 32  $\tilde S^3$'s for $\phi<1$ that vanish 
% when $\phi\to 1^-$. 
Therefore, 
  there is a one-to-one
correspondence between the sixteen $S^3\cup S^3$'s at $\phi>1$ and the  eight
 $\bigcup_{n=1}^4 \tilde S^3$ 
at $\phi<1$. As a consequence, on the $G_2$ orbifold, the two copies of 
 $S^3\cup S^3$ in Eq. (\ref{ss1}) at $x^{10}=0$ and $x^{10}=\pi R$ give 
rise to 4 $S^3$'s that
  undergo flop transitions. 

For completeness, we note that involutions involving  $z_3\to
e^{i\theta_3}\bar z_4$, $z_4\to
e^{i\theta_4}\bar z_3$, $z_5\to e^{i\theta_5}\bar z_5$ and their permutations 
do not give rise to vanishing cycles at $\phi=1$.

\subsection{A model with conifold transition}

The previous model is probably one of the simplest generalizations of
the M-theory phase transition considered in  \cite{Atiyah:2000zz} to
the case of a compact $G_2$ manifold and $SU(2)$ singularity.
But still the whole M-theory discussion takes place in the moduli
space of a single given $G_2$ space.
In \cite{Partouche:2001uq}, however, the possibility of a transition
into the moduli space of another smooth $G_2$ manifold was noticed. This
happens at the conifold point of the underlying CY manifold which is a
common boundary point in the complex structure moduli space of one CY
family and the (complexified) \Ka moduli space of another family. From the
four-dimensional point of view this corresponds to a new branch for
the Abelian/scalar sector of the effective field theory.
Therefore, we would now like to discuss in a compact model 
the appearance of this additional branch via {\em
  small resolution} of the CY. 

Let us  choose again a one
parameter sub-family of threefolds $\C_2$ within $\CP^4_{11222}[8]$,
defined  by 
\begin{equation}
p_2 = z_6^4(z_1^8 + z_2^8-2\phi z_1^4z_2^4)+(z_3^2-\phi z_6^2
z_2^4)^2+(z_4^2-\phi z_6^2 z_2^4)^2+(z_5^2-\phi z_6^2 z_2^4)^2=0
\label{def33} 
\end{equation}
with complex parameter $\phi$. 
The family $\C_2$ has singular members for $\phi=\pm 1$, $\phi=\pm i$
and $\phi=\pm i/\sqrt{2}$ where $p_2$ becomes
non-transverse. As in section \ref{model1}, the singular points lie in
 charts $z_2\neq0$, $z_6\neq 0$, so that we can rescale them to 1.
In anticipation of the antiholomorphic involution that we are going to
take, we shall be interested in the singularities that occur for real 
$\phi$,
\begin{equation}
\begin{array}{llll}
\mbox{for }\phi=+1 &~:~ & (i^k,1,\pm1,\pm1,\pm1,1)& ,(k=0,...,3)\\   
\mbox{for }\phi=-1 &~:~ & (i^k e^{i\pi/4},1,\pm1,\pm1,\pm1,1) &,(k=0,...,3)\; ,
\label{nodes2}
\end{array}
\end{equation}
where the $+/-$ signs are independent, giving rise to 32 singular
points on the CY in each case. The reason
why we added the monomials $z_A^2 z_6^2 z_2^4$, $(A=3,4,5)$, is
that they render the matrix of second derivatives of $p_2$
regular. The singularities on $\C_2$ are thus nodal points and $\phi$
sits at the conifold points in the complex structure moduli space of
$\C_2$. 

As in the previous model we proceed by restricting $\phi$ to real
values and consider the $G_2$ orbifold 
\begin{equation}
\G_2=(\C_2\times S^1)/\sigma\; ,
\end{equation}
where $\sigma=w\I$ is  defined as in Eq. (\ref{s}).
The special Lagrangian 3-cycle $\Sigma$ of $w$-invariant points in
$\C_2$, giving the $\sigma$-fixed points $\Sigma \times \{0,R\pi\}$,
is determined by 
\begin{equation}
x_6^4(x_1^8+x_2^8-2\phi x_1^4x_2^4)+(x_3^2-\phi x_6^2
x_2^4)^2+(x_4^2-\phi x_6^2 x_2^4)^2+(x_5^2-\phi x_6^2 x_2^4)^2=0 \; ,  
\label{fp2}
\end{equation}
where the unknowns are real and $x_{2}$, $x_6$ can be scaled to 1, thanks
to Eq. (\ref{forbid}). Solving for $x_1^4$, one obtains
\begin{equation}
x_1^4 =\phi \pm\sqrt{\phi^2-[1+
  (x_3^2-\phi)^2+(x_4^2-\phi)^2+(x_5^2-\phi)^2     ]}\; , 
\label{def3'}
\end{equation}
which implies $\phi\geq 1$. In the variables
\begin{equation}
u_1= x_1^4-\phi \;\; , \quad u_j=x_j^2-\phi \; , \quad
(j=3,4,5)\; ,
\label{maps}
\end{equation}
we find again a 3-sphere
\begin{equation}
u_1^2+u_3^2+u_4^2+u_5^2=\phi^2-1\; .
\label{s3} 
\end{equation}
Now, the four maps (\ref{maps}) from $x_i$ to $u_i$ are 
two-to-one and by the same reasoning as in the previous model one
easily shows that (for finite $\phi$) the $x$'s are either positive
definite or negative definite. Hence equation (\ref{s3}) gives rise to a
total of 16 disjoint copies of $S^3$.  
To summerize the situation, we have two distinct phases
along the real $\phi$ axis, 
$$
\begin{array}{ll}
\phi<1,~~~ & \Sigma={ \emptyset} \quad\mbox{(no fixed
  points)}\; , \\
\phi=1,~~~ & \Sigma=\{(\pm 1,1,\pm 1,\pm 1,\pm 1,1)\}\quad\mbox{(\ie 16
  points)}\; ,\\
\phi>1, ~~~& \Sigma= \bigcup_{n=1}^{16} S^3 \quad \mbox{of radii
  }\sqrt{\phi^2-1}\; , 
\end{array}
$$
where the $w$-invariant $S^3$'s that shrink to 16 of the 32 nodes in
Eq. (\ref{nodes2}) when $\phi\to 1^+$ are precisely the ones
involved in the conifold transitions of the underlying CY $\C_2$. 
In the next section we shall discuss the new branches of $G_2$
manifolds emanating from $\phi=\pm 1$ via  these CY conifold transitions.

\vskip .2in

\noindent
{\it Spectrum}
\vskip .05in 

Along the $\phi$ branches, the present orbifold $\G_2$ is similar to
$\G_1$. For completeness, we recall that the massless spectrum describes 
$b_3=89$ neutral chiral multiplets present for arbitrary $\phi$ from the
untwisted sector together with a pure $SU(2)$ super Yang-Mills
theory in non-Abelian $(\phi\gg1)$ or confining $(\phi\ll-1)$ phase for each 
of the 32 $S^3$ components of $\Sigma\times \{0,\pi R\}$. In total we have
\begin{equation}
\mbox{1 vector multiplet of }SU(2)^{32} \qquad \mbox{and} \qquad 
\mbox{89 neutral chiral multiplets .}
\end{equation}

\subsection{The extremal transition} 
\label{ext.trans}

Following \cite{Candelas:1990ug}, we would like to consider in this
section the conifold transition that occurs in $\C_2$ at
$\phi=\epsilon\equiv\pm 1$. We shall then generalize the results of
\cite{Partouche:2001uq} for the descendant extremal transitions
between $G_2$ spaces. 
 
At $\phi=\epsilon$, the non-transverse defining equation
(\ref{def33}) of $\C_2$ takes the form
\begin{eqnarray}
p_{\phi=\epsilon}= \mbox{Det}
\left(\begin{array}{rr}
P_{11}(z)&P_{12}(z) \\
-\bar{P}_{12}(z)&\bar{P}_{11}(z)\end{array}\right)\; ,
\label{det}
\end{eqnarray}
where 
\begin{equation}
\begin{array}{lll}
P_{11}(z)=z_6^2(z_1^4-\epsilon z_2^4)+i(z_3^2-\epsilon z_6^2 z_2^4)
&,& P_{12}(z)=(z_4^2-\epsilon z_6^2 z_2^4)-i(z_5^2-\epsilon z_6^2
z_2^4) \; ,
\end{array}
\label{pp}
\end{equation}
and $\bar{P}_{11}(z)$, $\bar{P}_{12}(z)$ are obtained by simply changing
$i$ to $-i$ in the coefficients.
The so called {\em small resolution} of this singular variety
consists
in replacing the 32 nodes given in Eq. (\ref{nodes2}) by 2-spheres
$S^2$. The result can then be written as an intersection of hypersurfaces 
defined by 
\begin{equation}
\left\{
\begin{array}{rl}
P_{11}(z)t_1+P_{12}(z)t_2=0& \\
-\bar{P}_{12}(z)t_1+\bar{P}_{11}(z)t_2=0& ,
\end{array}
\right.
\label{sr}
\end{equation}
where $(t_1,t_2)$ are projective coordinates parametrizing a
$\CP^1\equiv S^2$. In fact, since 
$(t_1,t_2)\neq (0,0)$, the determinant of the coefficients of
$t_{1,2}$, which is precisely $p_{\phi=\epsilon}$ in Eq. (\ref{det}),
must vanish
and the system in Eq. (\ref{sr}) is equivalent to 
\begin{equation}
\left\{
\begin{array}{ll}
& P_{11}(z)t_1+P_{12}(z)t_2=0\\
& p_{\phi=\epsilon}=0\; ,
\end{array} \label{eq1111}
\right.
\end{equation}
where the first equation determines $(t_1,t_2)$ in $\CP^1$. When the
volume of $\CP^1$ is sent to zero, the first 
equation is irrelevant and we recover the singular CY, while when
$\CP^1$ is of finite volume, the manifold $\C_2^{\epsilon }$ we
obtain is smooth and sits at  a
generic point of the family of complete intersection
CY's
\begin{equation}
\left[  \begin{array}{l||cc} \CP^4_{11222} &4&4 \\ 
\mathbb{CP}^1 &1&1 
   \end{array} 
\right]\, , 
\label{ncy}
\end{equation}
whose Hodge numbers are $h_{11}'=3$ and $h_{12}'=55$. \footnote{These Hodge
  numbers can be determined as follows. The Lefschetz hyperplane
  theorem implies that $h_{11}'$ is given by the dimension of the \Ka
  moduli space of the ambient space, therefore
  $h_{11}'=2+1$. Then, $h_{12}'$ is deduced from the relation $\chi' -
  \chi = 2 N$, where $\chi$ ($\chi'$) is the Euler characteristic of
  $\C_2$ ($\C_2^{\epsilon }$) and $N=32$ is the number of nodes.}   

We would like to see now if we can construct $G_2$ manifolds out of
these new CY branches.
We therefore consider products $\C_2^{\epsilon }\times S^1$ and look for an
antiholomorphic involution $w'$ acting on the whole $\C_2^{\epsilon }$ ambient
space $\CP_{11222}^4 \times\CP^1$. Since we know that $z_i$ should be
sent to $\bar{z}_i$, we need to extend the antiholomorphic action on
the coordinates $(t_1,t_2)$ of $\CP^1$. This action must be linear in
order to preserve the \Ka metric of $\CP^1$, so that there must be a 
$2\times 2$ matrix $M$ such that 
\begin{equation}
\pamatrix{t_1 \cr t_2}\to M\pamatrix{\bar{t}_1 \cr \bar{t}_2}\;,\qquad
MM^*=\lambda I \; , \qquad MM^\dag = \mu I\;, 
\end{equation}
where the two conditions on $M$ assure that the transformation is of
order two and preserves the \Ka metric of $\CP^1$.
As a result, the most general matrix $M$ (whose
determinant can be normalized without loss of generality) is
\begin{equation}
\pamatrix{a & iB \cr iB & \bar{a}}\; \mbox{ where } \;|a|^2+B^2=1\; ,\;
a\in \Co\;, \; B\in \R\qquad \mbox{or} \qquad \pamatrix{0&-1\cr 1&0}\;.
\end{equation}    
The first transformations always have fixed points in $\CP^1$ (as can
be seen by diagonalizing them, see \cite{Partouche:2001uq} for details), while
the second is freely acting (since, otherwise, $t_1=t_2=0$, which is
forbidden). Actually, among all these involutions, only the freely
acting one can be combined such that 
\begin{equation} 
w' \; : \quad z_i\to \bar{z}_i\quad (i=1,...,6)\quad , \qquad
t_1\to-\bar{t}_2\;, \;t_2\to \bar{t}_1 \; , 
\label{w'} 
\end{equation}
is a symmetry of $\C_2^{\epsilon}$, as can be seen from
\begin{eqnarray}
\left\{\begin{array}{c}
\phantom{-}P_{11}(z)y_1+P_{12}(z)y_2=0 \\
-\bar{P}_{12}(z)y_1+\bar{P}_{11}(z)y_2=0
\end{array} \right.\stackrel{w'}{\longrightarrow} &&
\left\{\begin{array}{c}
-P_{11}(\bar{z})\bar{y}_2+P_{12}(\bar{z})\bar{y}_1=0 \\
\phantom{-}\bar{P}_{12}(\bar{z})\bar{y}_2+\bar{P}_{11}(\bar{z})\bar{y}_1=0
\end{array}\; , \right.
\end{eqnarray}
after complex conjugation. We thus obtain
two new branches of smooth $G_2$ manifolds $(\C_2^{\epsilon }\times
S^1)/w'\I$ connected  
at $\phi=\pm1$ to the original ones $(\C_2\times S^1)/w\I$. 

Actually, as soon as we see that such branches exist, it is not a surprise to
observe that the orbifolds $(\C_2^{\epsilon }\times S^1)/w'\I$ cannot
have fixed points. In fact, such a fixed point set
would have been composed of two copies of a 
special Lagrangian 3-cycle $\Sigma^{\epsilon }$ on 
$\C_2^{\epsilon }$. Since on the $G_2$ 
branches built from $\C_2^{\epsilon }$ 
we vary \Ka classes of the CY, $vol(\CP^1)$ in our present model, 
the 3-cycle volume
$vol(\Sigma^{\epsilon })$ would remain constant when 
moving on this branch. As a result, by sending back $vol(\CP^1)\to 0$
towards the conical CY where $\Sigma^{\epsilon }=\Sigma$, we would find
$vol(\Sigma^{\epsilon })=0$. However, since $\C_2^{\epsilon }$ is
not singular, this implies that $\Sigma^{\epsilon }$ is empty.
In fact, for $\phi=-1$, this was ensured by the fact that
$p_{\phi=-1}$ in Eq. (\ref{eq1111}) has no solution when restricted to
real unknowns $x_i$'s. However, at $\phi=+1$, since $p_{\phi=+1}$ admits
real solutions, the antiholomorphic involution $w'$ has to act as
$t_1\to -\bar{t}_2\; , \; t_2\to \bar{t}_1$, so that it is freely
acting on $\C_2^{+1}$.  

\vskip .2in

\noindent
{\it Spectrum on $(\C_2^{\epsilon}\times S^1)/w'\I$}
\vskip .05in 

The spectrum on theses branches is
then immediately found. Since
$(dT\wedge d\bar{T})/|T|^2$, where $T=t_1/t_2$ is odd under $w'$, the third
2-cycle homology class of $\C_2^{\epsilon}$ is odd (together with the
two odd classes already present on $\C_2$). Consequently, the Betti
numbers take the values
\begin{equation}
b'_2=h_{11}^{+'}=0 \quad , \quad b'_3=1+h'_{12}+h_{11}^{-'}=59\; ,
\label{b'23}
\end{equation}
and the massless spectrum consists of $59$ chiral multiplets. 

\subsection{Physical interpretation of the new phases}

We would now like to interpret the phases described in the previous
section from a physical point of view. This means that instead of
considering geometrical moduli spaces of real dimensions $b_3$
($b_3'$), we
have instead to interpret the branches complexified with the eleven
dimensional supergravity 3-form $C$. The physical moduli space
coordinates are then the scalar components of the massless 
chiral fields. 

The extremal transition at $\phi=-1$ in characterized by 32 nodal
points on the underlying CY that are not fixed under the $\Z_2$
involution. These nodes are vanishing 2-spheres on $\C^{-1}_2$ whose
classes are odd under $w'$ (and equal in this example). This situation,
where $b_3$ varies and $b_2$ is constant,
is precisely one considered in \cite{Partouche:2001uq} where it was shown how
this describes a change of branch in the scalar potential, the
Abelian gauge group not being affected. At
$\phi=+1$, we saw that 16 of the nodes are mapped into each other under
$\Z_2$, while the 16 others are invariant. Again, the single
vanishing class of these nodes is odd. Hence this situation,
where fixed nodes arise, is a generalization of the previous case we
discuss now. 

Following \cite{Strominger:1995cz,Greene:1995hu}, 
suppose we have a CY $\C'$ where $N$ 2-cycles $\gamma_a$ $(a=1,...,N)$
shrink to zero size at some point in the \Ka moduli space. A priori
not linearly independent in homology, their classes satisfy
$R$ relations 
\begin{equation}
\alpha_1^r [\gamma_1] + \cdots +  \alpha_N^r [\gamma_N] = 0 \qquad
(r=1,...,R)\; ,
\end{equation}
with integer coefficients. When M-theory is compactified on $\C'\times
S^1$, membranes wrapped on these cycles give $N$ black hole
hypermultiplets charged under the $(N-R)$ independent $U(1)$'s
associated with the vanishing classes. In our case, $N=32$ and
$N-R=1$. If we denote by $(h_a,\tilde h_a)$ $(a=1,...,N)$ and $T^I$
$(I=1,...,N-R)$ the complex scalars of the hypermultiplets and vector
multiplets, respectively, $\N=2$ supersymmetry in four dimensions
implies the presence of the superpotential
\begin{equation}
{\cal W}=\sum_{I=1}^{N-R} \sum_{a=1}^{N} q_I^a T^I h_a \tilde h_a\; ,
\label{superpot}
\end{equation}
where $q_I^a$ is the charge of $(h_a,\tilde h_a)$ under the $I$-th
$U(1)$.  Since the antiholomorphic involution is an isometry, it maps
the set of nodes into itself. We can therefore define $N=N'+N''$,
where $N'$ is even and counts the number of non-invariant nodes, while
$N''$ is the number of fixed nodes. We order the nodal points so
that the vanishing 2-cycles on $\C'$ satisfy $\gamma_{b}\to
-\gamma_{{b}+N'/2}$ $(b=1,...,N'/2)$ and $\gamma_{c}\to -\gamma_{c}$
$(c=N'+1,...,N'+N'')$, the minus sign being a consequence of the fact
that these cycles are calibrated and thus holomorphic while the
involution is antiholomorphic. In these conventions we have 
\begin{equation}
\begin{array}{lll}
h_{b}(z,T,x^{10})=\tilde h_{b+N'/2}(\bar z,-1/\bar T,-x^{10})& ,&
\tilde h_{b}(z,T,x^{10})=h_{b+N'/2}(\bar z,-1/\bar T,-x^{10})\\ 
h_{c}(z,T,x^{10})=\tilde h_{c}(\bar z,-1/\bar T,-x^{10}) &,&
\end{array}
\end{equation}
so that the sums $H_b=h_b+\tilde h_{b+N'/2}$, $\tilde H_b=\tilde
h_b+h_{b+N'/2}$ and $H_c=h_c+\tilde h_{c}$ are even under the
involution $\sigma'=w'\I$ and become the scalar components of $\N=1$ chiral
multiplets, while the differences are projected out. The
superpotential (\ref{superpot}) now becomes 
\begin{equation}
{\cal W}= \sum_{I=1}^{N-R}\left( \sum_{b=1}^{N'/2} q_I^b T^I H_b
  \tilde H_b +  \frac{1}{2}\sum_{c=N'+1}^{N'+N''}q_I^c T^I H_c
  H_c\right) \;,
\end{equation}
with classical diagonal kinetic terms. The scalar potential has 
different phases. When the scalars $T^I$ have non-vanishing vacuum
expectation values, all the black hole chiral multiplets are
massive. This branch describes the model on the manifold $(\C'\times
S^1)/{w'\I}$ (with $b'_3=59$ chiral multiplets in our example). In the reverse
situation where the black holes condense, 
the  $T^I$ fields become massive. The massless spectrum contains in
addition to the $b'_3$ chiral multiplets the $N'/2$ pairs $(H_b,\tilde
H_b)$ and $N''$ fields $H_c$, from which we substract $(N-R)$ complex 
degrees of
freedom arising from the vanishing F-term conditions associated to the
$T^I$'s and another $(N-R)$ from the massive $T^I$'s themselves. On
this branch, the massless spectrum is then composed of an equal number
$b_2$ (equal to 0 in our example)
 of $U(1)$ factors and $b_3$ chiral multiplets given
by\footnote{We hope the reader will not be confused by the fact that
  we reversed the notations of ref. \cite{Partouche:2001uq} for the primed and not
  primed Betti numbers.}
\begin{equation}
b_2=b_2'\quad \mbox{and} \quad  b_3=b_3' +2R-N\; ,
\label{newBetti}
\end{equation}
in concordance with the model of the present section, $b_3=59+2\cdot
31-32=89$ for $\phi=1$ ($\phi=-1$), where $N'=16$, $N''=16$, ($N'=32$,
$N''=0$), respectively. 
We see that actually only the total number of nodes $N$
is relevant in the transition. However, the pairs of chiral fields
associated to  $H_b$ and $\tilde H_b$ form an $\N=2$ sector in the
theory, while the fields $H_c$ are truly part of $\N=1$
multiplets. From an M-theory point of view, the former are associated
with  membranes wrapped on copies of $\CP^1$ identified two by
two on the orbifold, so that they are unoriented, 
while the latter are associated with  unoriented membranes wrapped
on $\RP^2$.     

Having extended the validity of Eq. (\ref{newBetti}) 
for extremal transitions at constant $b_2$ to the case where
nodal points are fixed under the involution, we would like to note
that there is no similar extension for the case where $b_2$ varies
and describes a Higgs mechanism of the Abelian gauge group. The reason is
that the class of a fixed nodal point satisfies
$[\gamma_c]\to -[\gamma_c]$ under the antiholomorphic involution and
that the Higgs effect described in \cite{Partouche:2001uq}
arises when a vanishing homology
class is even.

\section{Models with adjoint matter  \label{MM}}

\subsection{A model based on {$(\CP^4_{11222}[8] \times S^1)/ \Z_2$}}

We have seen before that the compactification of M-theory on $G_2$-holonomy
spaces with associative cycles $\Sigma$ of 
$A_{N-1}$ singularities describes $SU(N)$ gauge
theories with $b_1(\Sigma)$ Wilson lines in chiral multiplets. In Section 2,
we presented models with $b_1(\Sigma)=0$. Here, we will discuss a simple
realization of a gauge theory with matter in the adjoint arising from 
$b_1(\Sigma)>0$. For this,  
let us consider another family of CY's $\C_3$ in $\CP^4_{11222}$,
similar to the one considered in Section 2.1. The Hodge numbers
are still $h_{11}=2$ and $h_{12}=86$ and the defining polynomial is now 
\begin{equation}
p_3 = z_6^4(z_1^8 + z_2^8)+z_3^4+z_4^4+z_5^4-2\theta z_4^2z_5^2=0\; ,
\label{def333} 
\end{equation}
where  $\theta$ is a complex parameter. 
Thanks to Eq. (\ref{forbid}), $\C_3$ happens to be singular when
$z_3=z_6=0$, $\theta=\pm1$ and $z_4^2=\theta z_5^2\neq 0$. As a
result, we can set $z_5$ to 1 by a  $\Cs_1$ rescaling and obtain the
sets of singular points,
\begin{equation}
\begin{array}{llll}
\mbox{for }\theta=+1 &~:~ & (z_1,z_2,0,\pm1,1,0)& ,\\   
\mbox{for }\theta=-1 &~:~ & (z_1,z_2,0,\pm i,1,0) &,
\label{p1sing}
\end{array}
\end{equation}
where $(z_1,z_2)\neq (0,0)$ are arbitrary and modded out by the
$\Cs_2$ action, so that they parametrize a $\CP^1$. As
a consequence, two non-intersecting 
 genus 0 curves of singularities occur on $\C_3$
when $\theta=\pm1$. 

To construct $G_2$ manifolds out of this family of CY's, we consider
the involution $\sigma= w\I$ in Eq. (\ref{s}) and define
\begin{equation}
\G_3=(\C_3\times S^1)/\sigma\; ,
\end{equation}
where the complex structure parameter $\theta$ is now restricted to be real
for $w$ to be an isometry of $\C_3$. Before complexification with the
M-theory 3-form $C$, the $G_2$ moduli space is again split in three
disconnected pieces, $\theta<1$, $-1<\theta<1$ and $\theta>1$.   

The fixed point set on $\G_3$ is again two copies 
at $x^{10}=0$ and $x^{10}=\pi R$ of the special
Lagrangian 3-cycle $\Sigma$ fixed by $w$ on $\C_3$. Its equation
reduces to the polynomial $p_3$ for real unknown $x_i$,
\begin{equation}
x_6^4(x_1^8 + x_2^8)+x_3^4+x_4^4+x_5^4-2\theta x_4^2x_5^2=0\; .
\label{fp3} 
\end{equation}
From Eq. (\ref{forbid}), the factor multiplying $x_6^4$ is always
strictly positive. Therefore, $x_5=0$ would also imply $x_3=x_4=x_6=0$
which is forbidden, so that we can rescale $x_5$ to 1. Now, we observe
that $(x_1,x_2)\neq(0,0)$ modded out by the real rescaling reminiscent
to $\Cs_2$ parametrize a full $\RP^1\equiv S^1$. Let us now fix a
point $(x_1,x_2)$ in this $S^1$, and then define the one-to-one change
of variables
\begin{equation}
u_6 = x_6^2 \sqrt{x_1^8+x_2^8}\, \mbox{sign}(x_6) \; , \quad u_3=x_3^2\,
\mbox{sign}(x_3)\; , 
\end{equation}
and write Eq. (\ref{fp3}) as
\begin{equation}
u_6^2+u_3^2=- (x_4^4-2\theta x_4^2 +1)\; .
\label{s2}
\end{equation}
This equation has solutions when the r.h.s. is positive \ie when
\begin{eqnarray}
&&0<(\theta-\sqrt{\theta^2-1})^{1/2}
\leq x_4\leq (\theta+\sqrt{\theta^2-1})^{1/2}\\  &\mbox{ or }&
\phantom{0}-(\theta+\sqrt{\theta^2-1})^{1/2}
\leq x_4\leq -(\theta-\sqrt{\theta^2-1})^{1/2}<0\; , 
\label{x4}
\end{eqnarray}
implying $\theta\geq 1$. 
As a conclusion, when choosing $x_4$ in these segments,
Eq. (\ref{s2}) describes a circle parametrized by $u_{6,3}$ whose
radius vanishes at the boundaries of the $x_4$ intervals. We thus
obtain two 2-spheres at each point of the $S^1$ parametrized by
$x_{1,2}$. Since the $S^2$'s are disjoint,  this results in
two copies of $S^2\times S^1$. To summerize, the fixed point set
according to the values of $\theta$ is
$$
\begin{array}{ll}
\theta<1,~~~ & \Sigma={ \emptyset} \quad\mbox{(no fixed
  points)}\; , \\
\theta=1,~~~ & \Sigma=\{(x_1,x_2,0,\pm1,1,0)\}\quad \mbox{{\em
  \ie}}\quad S^1\cup S^1\; ,\\
\theta>1, ~~~& \Sigma= (S^2\times S^1)\cup (S^2\times S^1) \; ,
\end{array}
$$
where the radius of the $S^2$'s is of order $\sqrt{\theta-1}$ when
$\theta\to 1^+$. Note that the complex deformation of $\C_3$
can determine the volume of these 2-spheres since
they are not holomorphic on the CY. Also, at $\theta=1$,
the $S^1$'s are the equators of the
2-spheres of singularities in Eq. (\ref{p1sing}). We thus have a
transition between two distinct phases $\theta<1$ and
$\theta>1$.
As in Section 2.4, there could also be extremal
transitions at $\theta=\pm1$ giving rise to new branches of $G_2$
manifolds. However, we don't consider this situation in this work. 
 
\vskip .2in

\noindent
{\it Spectrum for $\theta<1$}
\vskip .05in 

On this branch, the orbifold is smooth and the spectrum is thus
simply determined by the Betti numbers from Eq. (\ref{b23}) to give
the same result we had for $\G_1$ when $\phi<1$,
$$
b_2=0 \qquad \mbox{ and } \qquad b_3=1+86+2=89\; ,
\label{q'}
$$ 
\ie $89$ chiral multiplets with no gauge group.  

\vskip .2in

\noindent
{\it Spectrum for $\theta>1$}
\vskip .05in 

In addition to the previous spectrum, one now has
to take into account the states that sit at each connected  component
of  the fixed point set 
$\Sigma\times \{0,\pi R\}$.  As in the case $\phi>1$ for $\G_1$, for each
$\Sigma_p=S^2\times S^1$ $(p=1,2)$,  they fill a
$\N=1$ multiplet in 7 dimensions in the adjoint of $SU(2)$ 
compactified down to 4 dimensions on $\Sigma_p$ 
giving rise to a $\N=1$
vector multiplet. In addition, we get in general  $b_1(\Sigma_p)$ (equal to 1
in our case)  
real scalars 
in the adjoint of $SU(2)$, arising
from the flux of the 7 dimensional gauge boson on the 1-cycles of $\Sigma_p$.
Notice that the vacuum expectation value
of these scalars are zero since otherwise they would break the gauge
group. Actually, they are  combined with the volume of the 
vanishing 2-sphere at the $A_1$ singularity times the radius $r_\alpha$ 
$(\alpha=1,...,b_1(\Sigma_p))$ to give $b_1(\Sigma_p)$ complex scalars in the 
adjoint of $SU(2)$. As a result, the total spectrum in the present case
consists in 
\begin{eqnarray} 
&\mbox{ 1 vector and 1 chiral multiplet in the adjoint  of
  $SU(2)^4$}&\nonumber  \\ &
\mbox{and 89 neutral 
chiral multiplets .}&
\end{eqnarray}

Since $b_1(\Sigma_p)>0$, we can consider now the  desingularization of $\G_3$
while keeping the $G_2$ holonomy. To this end we have to blow up the 
2-spheres vanishing at each $A_1$ singularity. This amounts to give a
mass to the  membranes wrapped on them and corresponds physically to move in
the Coulomb branch  $SU(2)^4\to U(1)^4$. The chiral multiplets associated to
these 2-cycles get a   vacuum expectation value and become massive. As a
result, in addition to the ``untwisted''  $b_2=0$ vector  multiplets
and $b_3=89$ chiral 
multiplets, we now have a ``twisted sector'' consisting of 
$2P$ vector multiplets and $2 \sum_{p=1}^P b_1(\Sigma_p)$ 
neutral chiral multiplets (where  
$P=2$ in the present model). In general, 
 in the Coulomb branch of each of the $2P$ $SU(2)$ coupled to
 $b_1(\Sigma_p)$  
chiral multiplets, we have
\begin{eqnarray}
&&\phantom{\mbox{and}~~~~}b_2=h_{11}^++2P ~~~~~~~~~~~~~~~~~~~~~~~~~~~~
 \mbox{vector multiplets} 
\nonumber \\
&&\mbox{and}~~~~
b_3=1+h_{12}+h_{11}^- + 2\sum_{p=1}^P b_1(\Sigma_p)~~~~   
\mbox{chiral multiplets .}\label{JjJ} 
\end{eqnarray}
 Note that this result was 
conjectured in \cite{Kachru:2001je} for the case $P=1$
by considering the dual description in type IIA in the presence
of D6-branes and O6-planes\footnote{Eq. {(\ref{JjJ})} 
was also known to the authors of \cite{Partouche:2001uq}.}.
In our example, since
we have $P=2$ disconnected components $S^2\times S^1$ in $\Sigma$,
each of them with $b_1=1$
the full massless theory 
turns out to be a $U(1)^4$ gauge theory and $89+4=93$ neutral
chiral multiplets. The formula (\ref{JjJ}) is  actually 
an extension of the case 
considered by Joyce in the desingularization of orbifolds $T^7/\Gamma$, where
the fixed points are $2P$ copies of non-intersecting $T^3$'s
\cite{Joyce2,Joyce1}.   
 
Finally, let us note that in this example, the twisted sector 
 realizes a classical
 pure ${\cal N}=2$ 
$SU(2)$ super Yang-Mills theory in M-theory, however coupled to 
the ${\cal N}=1$  chiral multiplet associated to $\theta$.

\subsection{Yet another model with {$S^2 \times S^1$} flops}
 
There are other CY's considered in the literature containing
 special Lagrangian 
3-cycles with $b_1(\Sigma)>0$. As an example, model III of  
\cite{Kachru:2000an}
is defined by the polynomial 
\begin{equation}
p_4 = z_6^4(z_1^8 + z_2^8-2\phi z_1^4z_2^4)+z_3^4+z_7^2(z_4^4+z_5^4)=0\; ,
\label{defff} 
\end{equation}
in  $\CP^4_{11222}$, where the  $z_7$ coordinate arises by orbifolding 
our model $\C_1$ by the $\Z_2$ action $(z_4,z_5)\to (-z_4,-z_5)$ and
blowing up the singularity. Therefore,
there is a third $\Cs$ action acting as $(z_4,z_5,z_7)\to (\lambda z_4,
\lambda z_5, \lambda^{-2}z_7)$ and the excluded set of the ambient
toric variety is $(z_1,z_2)\neq(0,0)$, $(z_4,z_5)\neq(0,0)$ and
$(z_3,z_6)\neq(0,0)$.  
The Hodge numbers of the CY are 
$h_{11}=3$ and $h_{12}=55$. 
With the involution (\ref{s}) extended by $z_7 \to \bar{z}_7$,
the analysis of this model is analogous 
to the model based on $\C_1$. The spectrum for $\phi<1$ turns out to be 
$$
b_2=0 \qquad \mbox{ and } \qquad b_3=1+55+3=59\; ,
\label{q'q}
$$ 
\ie $59$ chiral multiplets with no gauge group. For $\phi>1$, 
 Eq. (\ref{Sph}) gets modified to 
\begin{equation}
u_1^2+u_3^2+u_7^2=\phi^2-1\; ,  \label{Sphh}
\end{equation}
where $u_7=x_7\sqrt{x_4^4+x_5^4}$. Here, $(x_4,x_5)$ parametrize an 
$\RP^1\equiv S^1$ moded  out by $(z_4,z_5)\to (-z_4,-z_5)$ giving rise to 
another 
$S^1$ of half radius,
 and Eq. (\ref{Sphh}) defines an $S^2$ in
$u$-variables. Therefore,   
the $S^3$'s of $\C_1$ are replaced by $S^2\times
S^1$'s in this case, so that
$\Sigma=S^2\times S^1 \cup S^2\times S^1$ has $P=2$ components 
$\Sigma_p=S^2\times S^1$
with $b_1(\Sigma_p)=1$.  
As in the previous section,  the full spectrum in this branch is 
\begin{eqnarray}
&\mbox{ 1 vector and 1 chiral multiplets in the adjoint  of
  $SU(2)^4$}&\nonumber  \\ &
\mbox{and 59 neutral chiral multiplets .}&
\end{eqnarray}
In addition, the $\tilde S^3$'s described in Eq. (\ref{tt}) for 
$\phi <1$ become
similarly $\tilde S^2\times S^1$'s given by 
\begin{equation}
 2(\phi +1) U_1^2+u_3^2+u_7^2=2(1-\phi)V_1^2\; . 
\end{equation}
%with $U_1$ and $V_1\to 1$ defined as in the former case.
As a result, this should describe   $S^2\times
S^1$ flops on the $G_2$ orbifold, 
where only the 2-spheres vanish.

 Finally, the Coulomb branch occurs as before 
when the orbifold fixed points for $\phi>1$ are desingularized. The
 resulting spectrum is then  
$b_2=0+4$ vector multiplets and $b_3=59+4=63$ chiral multiplets.

\vskip .5in

\noindent
This work is partially supported by the RTN contracts HPRN-CT-2000-00122 and 
 HPRN-CT-2000-00131 and the $\Gamma\Gamma$ET grant E$\Lambda$/71.


\begin{thebibliography}{999}

\bibitem{Sen:1997kz}
A.~Sen,
``A note on enhanced gauge symmetries in M- and string theory,''
JHEP{\bf 9709}, 001 (1997), hep-th/9707123.

\bibitem{FKPZ}
S.~Ferrara, A.~Kehagias, H.~Partouche and A.~Zaffaroni,
``Membranes and fivebranes with lower supersymmetry and their AdS  
supergravity duals,''
Phys.\ Lett.\ B {\bf 431}, 42 (1998), hep-th/9803109.

\bibitem{Papadopoulos:1995da}
G.~Papadopoulos and P.~K.~Townsend,
``Compactification of D = 11 supergravity on spaces of exceptional holonomy,''
Phys.\ Lett.\ B {\bf 357}, 300 (1995),
hep-th/9506150.


\bibitem{Partouche:2001uq}
H.~Partouche and B.~Pioline,
``Rolling among G(2) vacua,''
JHEP{\bf 0103}, 005 (2001),
hep-th/0011130.

\bibitem{Acharya:2000gb}
B.~S.~Acharya,
``On realising N = 1 super Yang-Mills in M theory,''
hep-th/0011089.

\bibitem{Atiyah:2000zz}
M.~Atiyah, J.~Maldacena and C.~Vafa,
``An M-theory flop as a large n duality,''
hep-th/0011256.

\bibitem{Acharya:2001dz}
B.~Acharya and C.~Vafa,
``On domain walls of N = 1 supersymmetric Yang-Mills in four dimensions,''
hep-th/0103011.

\bibitem{Joyce2}D.D. Joyce, 
``Compact Riemannian 7-manifolds with $G_2$ holonomy, II", 
J. Diff. Geom. {\bf 43} (1996) 329.

\bibitem{Candelas:1990ug}
P.~Candelas, P.~S.~Green and T.~Hubsch,
%``Rolling Among Calabi-Yau Vacua,''
Nucl.\ Phys.\ B {\bf 330}, 49 (1990).

\bibitem{Strominger:1995cz}
A.~Strominger,
%``Massless black holes and conifolds in string theory,''
Nucl.\ Phys.\ B {\bf 451}, 96 (1995), hep-th/9504090.

\bibitem{Greene:1995hu}
B.~R.~Greene, D.~R.~Morrison and A.~Strominger,
%``Black hole condensation and the unification of string vacua,''
Nucl.\ Phys.\ B {\bf 451}, 109 (1995), hep-th/9504145.

\bibitem{Pap}
J. Gutowski, G. Papadopoulos, ``Moduli spaces and brane solitons for 
M-theory compactifications on holonomy $G_2$ manifolds", hep-th/0104105.

\bibitem{Gomis:2001vk}
J.~Gomis,
``D-branes, holonomy and M-theory,''
hep-th/0103115.


\bibitem{Edelstein:2001pu}
J.~D.~Edelstein and C.~Nunez,
``D6 branes and M-theory geometrical transitions from gauged  supergravity,''
hep-th/0103167.

\bibitem{Kachru:2001je}
S.~Kachru and J.~McGreevy,
``M-theory on manifolds of $G_2$ holonomy and type IIA orientifolds,''
hep-th/0103223.
 
\bibitem{Cachazo:2001}
F.~Cachazo, K.~Intriligator and C.~Vafa, ``A large N duality via a
geometric transition'', hep-th/0103067.

\bibitem{Edelstein:2001b}
J.~D.~Edelstein, K.~Oh and R.~Tatar,
``Orientifold, geometrical transitions and large N duality for $SO$/$Sp$
gauge theories'', hep-th/0104037. 

\bibitem{Harvey:1999a}
J.~A.~Harvey and G.~Moore, ``Superpotentials and membrane instantons'',
hep-th/9907026. 

\bibitem{Joyce1}
D.D. Joyce, 
``Compact Riemannian 7-manifolds with $G_2$ Holonomy, I", 
J. Diff. Geom. {\bf 43} (1996) 291.

\bibitem{Joyce11}  D.D. Joyce, {\it private communication.}

\bibitem{Salamon:1989}
S.M.~Salamon, ``Riemannian geometry and holonomy groups'', Pitman
Res.~Notes in Math.\ 201, Longman, Harlow 1989.

%\bibitem{Seiberg:1994rs}
%N.~Seiberg and E.~Witten,
%%``Electric - magnetic duality, monopole condensation, and
%%confinement in N=2 supersymmetric Yang-Mills theory,'' 
%Nucl.\ Phys.\ B {\bf 426}, 19 (1994);
%Erratum-ibid.\ B {\bf 430}, 19 (1994), 
%hep-th/9407087.


\bibitem{Kachru:2000an}
S.~Kachru, S.~Katz, A.~E.~Lawrence and J.~McGreevy,
``Mirror symmetry for open strings,''
Phys.\ Rev.\ D {\bf 62}, 126005 (2000),
hep-th/0006047.


\end{thebibliography}
\end{document}